\documentclass[twocolumn,showpacs,aps,prl]{revtex4}

\newcommand{\bk}{{\bf k}}

\newcommand{\bP}{{\bf P}}
\newcommand{\br}{{\bf r}}

\newcommand{\bs}{{\bf s}}

\newcommand{\beq}{\begin{eqnarray}}
\newcommand{\eeq}{\end{eqnarray}}
\newcommand{\beqq}{\begin{eqnarray*}}
\newcommand{\eeqq}{\end{eqnarray*}}

\usepackage{graphicx}
\usepackage{dcolumn}
\usepackage{bm}
\usepackage{color}

\begin{document}

\begin{titlepage}

\title{Parity-breaking phases of spin-orbit-coupled metals with gyrotropic, ferroelectric and multipolar orders}
\author{Liang Fu}
\affiliation{Department of Physics, Massachusetts Institute of Technology,
Cambridge, MA 02139, USA}

\begin{abstract}
We study Fermi liquid instabilities in spin-orbit-coupled metals with inversion symmetry. 
By introducing a canonical basis for the doubly degenerate Bloch bands in momentum space, we 
derive the general form of Landau interaction functions.  A variety of time-reversal-invariant, parity-breaking phases is found, whose Fermi surface is spontaneously deformed and spin-split. In terms of symmetry, these phases possess gyrotropic, ferroelectric and multipolar orders. The ferroelectric and multipolar phases are accompanied by structural distortions, from which the electronic orders can be identified. The gyrotropic phase exhibits a unique nonlinear optical property. We identify correlated electron materials that exhibit these parity-breaking phases, including LiOsO$_3$ and Cd$_2$Re$_2$O$_7$.    
\end{abstract}


\maketitle

\draft

\vspace{2mm}

\end{titlepage}

Novel physics from strong spin-orbit coupling in quantum materials is currently attracting widespread interest across many disciplines in condensed matter physics.  
In particular, there is now an intensive investigation of the interplay between spin-orbit coupling and electron correlation in $d$-orbital and $f$-orbital systems\cite{balents, sun, senthil}.  
The majority of studies have been focused on correlated band or Mott insulators, whereas spin-orbit coupling in correlated metals has received less attention. 
It is well-known that spin-orbit coupling in metals without inversion symmetry generates spin-split energy bands and spin-polarized Fermi surfaces\cite{balents-fl}. 
This has interesting consequences in the presence of electron-electron interactions\cite{hirsch, berg, maslov, rashba, soc-nematic, ruhman, potter}.      
In contrast, in metals with inversion and time-reversal symmetry,  Bloch bands are doubly degenerate everywhere in momentum space. 
The effect of spin-orbit coupling is more subtle: it leads to spin-orbit-entangled Bloch wavefunctions, 
which have different spin-polarizations on different atomic orbitals\cite{fukane, zunger}.  
For this reason, the importance of spin-orbit coupling in inversion-symmetric materials can be easily overlooked. 
 
In this Letter, we explore the consequences of having both strong spin-orbit coupling and electron interaction in metals with inversion symmetry. 
By generalizing Landau's Fermi liquid theory to spin-orbit-coupled metals, 
we theoretically predict a variety of new ordered phases resulting from Pomeranchuk type instabilities in the spin channel\cite{pomeranchuk}, which  
spontaneously break inversion symmetry. These phases can be regarded as 
new examples of electronic liquid crystals\cite{nematic}, which preserve the translational invariance and break the point group symmetry of the lattice. 
Importantly, because of the spin-orbit coupling, these phases exhibit spin-split Fermi surfaces with characteristic spin textures, 
and the onset of electronic parity-breaking orders is generally accompanied by structural changes. 
We focus on three different parity-breaking phases having the symmetry of ferroelectric, a multipolar and an isotropic gyrotropic liquid respectively, and 
 identify their realizations in  
correlated electron materials.  
  
Landau's Fermi liquid theory of metals starts from Bloch states on the Fermi surface. 
In the presence of spin-orbit coupling, Bloch states are not spin eigenstates, but remain doubly degenerate at every $\bf k$  in systems with both time-reversal ($T$) and inversion ($P$) symmetry\cite{fukane}. To develop Fermi liquid theory of such spin-orbit-coupled systems, we must first choose a basis 
$\{ |\psi_{\bk, 1} \rangle, |\psi_{\bk, 2}\rangle \}$ 
for the degenerate bands over the entire Fermi surface.   
As observed by Blount long ago\cite{blount}, due the absence of spin conservation, 
the choice of basis is not unique: an arbitrary $U(2)$ rotation on the doublet at every $\bk$ 
produces a new basis that appears to be as good as the old one. This leads to significant complications, as the form of Landau energy functional is basis dependent.  

In this work, we introduce a canonical basis that we call ``manifestly covariant Bloch basis'' (MCBB). This basis is defined {\it universally and uniquely}  
by demanding the Bloch wavefunctions at $\br=0$---a  two-component spinor---to be fully spin-polarized along a global spin-quantization axis: 
\beq
\psi_{\bk, 1}(\br=0) &=& u_\bk |\uparrow \rangle, \nonumber\\
\psi_{\bk, 2}(\br=0) &=&  v_\bk |\downarrow \rangle. \label{def}
\eeq 
where $u_\bk$ and $v_\bk$ are {\it real} and {\it positive}; $\uparrow, \downarrow$ labels electron's spin.  Importantly, the origin of real space coordinate $\br=0$  is chosen to be the center of 
point group symmetries of the crystal; and the condition (\ref{def}) is imposed on Bloch states on the entire Fermi surface. 

The MCBB can be explicitly constructed by starting from an arbitrary basis $\{ |\phi_{\bk,1}\rangle, |\phi_{\bk, 2}\rangle \}$. Because of time-reversal ($T$) and inversion ($P$) symmetry, the two members form a Kramers doublet under the combined operation $PT$\cite{fukane}, so that with a proper choice of phase, we have $|\phi_{\bk,2}\rangle = PT |\phi_{\bk, 1}\rangle$. Therefore, the corresponding spinors defined by Bloch wavefunctions at the inversion center form a Kramers doublet under $T$: $\phi_{\bk,2}(\br=0) = T\phi_{\bf,1}(\br=0)$, and thus are orthogonal.     
This orthogonality condition guarantees one can perform a $U(2)$ transformation on $\{ |\phi_{\bk,1}\rangle, |\phi_{\bk, 2}\rangle \}$ to obtain a new basis satisfying (\ref{def}), or equivalently MCBB.    

The advantage of MCBB lies in its remarkably simple transformation property under point group symmetries, which act on both the electron's spatial coordinate and spin. 
For a generic choice of basis, a symmetry action $G$ will map Bloch states at $\bk$ into those at $G\bk$ (or the star of $\bk$) up to a complicated, $\bk$-dependent $U(2)$ basis transformation\cite{blount}. 
In contrast, the defining property  (\ref{def}) guarantees that $G$ maps the MCBB  $|\psi_{\bk, \alpha}\rangle$ 
at $\bk$ directly to its partner at $G\bk$,  
\beq
G:   |\psi_{\bk, \alpha} \rangle \rightarrow U_{\alpha \beta}(G) | \psi_{G\bk, \beta}\rangle \label{bsym}
\eeq
where $U(G)$ is the $SU(2)$ matrix representation of $G$. 
 Furthermore,  the MCBB at $\pm \bk$ are related by  time reversal symmetry in the same way as spin eigenstates:  
\beq
T|\psi_{\bk, \alpha}\rangle = \epsilon_{\alpha, \beta} |\psi_{-\bk, \beta}\rangle. \label{tsym}
\eeq
Eq.(\ref{bsym}) and (\ref{tsym}) show that the two members of the MCBB $\alpha=1,2$ 
 transform identically as spin up and down under symmetry operations. Therefore, for the simplicity of presentation, we will refer to the $\alpha$   index of MCBB 
 as spin, with the understanding that $|\psi_{\bk, \alpha}\rangle$ are not spin eigenstates. 
MCBB provides the starting point for our Fermi liquid theory, and is expected to have wide applications in spin-orbit-coupled systems in general.

Fermi liquid theory relates the change of energy $\delta E$ to the change in the distribution function of Bloch quasi-particles up to second order. 
The distribution function is a $2\times 2$ Hermitian matrix in spin space, which we write as $n_{\alpha\beta}(\bk)$ in MCBB, i.e., 
$n_{\alpha\beta}(\bk) =  \langle c^\dagger_{\bk,\alpha} c_{\bk, \beta}\rangle$. $\delta E$ is then a quadratic functional of $n_{\alpha\beta}(\bk)$, where $\bk$ is near the Fermi surface. 
For spin-orbit-coupled systems, we find it convenient to decompose $n_{\alpha \beta}(\bk)$ in terms of the density and spin distribution function: 
\beq
n_{\alpha \beta}(\bk) \equiv n(\bk) \delta_{\alpha \beta} + {\bf s}(\bk) \cdot {\vec \sigma}_{\alpha \beta}. 
\eeq
Based on symmetry considerations, we now relate $\delta E$ to the change in the density and spin distribution function. First, note the transformation property of 
$n(\bk)$ and $\bs(\bk)$ under time reversal and inversion, 
\beq
&T&:  n(\bk) \rightarrow n(-\bk), \; \bs(\bk) \rightarrow -\bs (-\bk) \nonumber\\
&P&: n(\bk) \rightarrow n(-\bk), \; \bs(\bk) \rightarrow \bs (-\bk). 
\eeq 
It follows that when both symmetries are present, $\delta E$ consists of density-density interaction and spin-spin interaction, taking the form of  
\beq
\delta E &=& \sum_\bk \epsilon_\bk \delta n(\bk) + \sum_{\bk, \bk'}  F^n(\bk, \bk') \delta n(\bk)  \delta n(\bk') \nonumber \\
&+&  \sum_{\bk, \bk'}  F^s_{ij} (\bk, \bk')   s_i(\bk) s_j(\bk'). \label{E}
\eeq
 
The invariance of $\delta E$ under crystal symmetry transformations further constrains the momentum dependence of the interaction functions $F^n(\bk, \bk')$ and $F^s(\bk, \bk')$.  
Unlike spin-rotationally-invariant systems where the spin interaction is isotropic in spin space ($F^s_{ij} \propto \delta_{ij}$),  
 both $F^s(\bk, \bk')$ and $F^s_{ij}(\bk, \bk')$ in spin-orbit-coupled systems are constrained 
 by crystal symmetries acting on electron's coordinate and spin in combination.  It follows from the symmetry property of MCBB (\ref{bsym})  that     
under a crystal symmetry operation $G$, $n(\bk)$ and ${\bf s}(\bk)$ transform as a scalar and a vector field respectively 
\beq
&G:& n(\bk) \rightarrow n(G\bk) \nonumber \\
&& s_i (\bk) \rightarrow G_{ij} s_j (G\bk).  \label{nssym}
\eeq    
where $G_{ij}$ is the $SO(3)$ matrix representation of $G$. 
Hence, $F^n(\bk, \bk')$ and $F_{ij}^s(\bk, \bk')$ transform as a scalar field and a rank-two tensor field respectively:  
\beq
&G: & F^n(\bk, \bk') \rightarrow F^n(G\bk, G\bk')  \nonumber \\
&& F^s_{ij}(\bk, \bk') \rightarrow G_{i i'} G_{j j'} F^s_{i'j'}(G\bk, G\bk'). \label{fsym}
\eeq 
Eq.(\ref{E}) and (\ref{fsym}) give the general form of the energy functional of Fermi liquids in spin-orbit-coupled systems.     
The novelty here lies in the spin interaction, 
which is anisotropic in spin space and spin-momentum locked. 
We will now explore consequences of spin interactions in spin-orbit-coupled Fermi liquids. 

To proceed, we write the spin interaction in a separable form given by products of basis functions of $\bk$ and of $\bk'$:
\beq
\delta E_{\rm spin} &\equiv& \sum_{\bk, \bk'} F^s_{ij}(\bk, \bk') s_i(\bk) s_j(\bk')  \nonumber \\
&=&\sum_\eta \sum_{\bk, \bk'} F_\eta \phi_{\eta} (\bk, \bs(\bk)) \phi_\eta (\bk', \bs(\bk')). \label{fs}
\eeq 
Naturally, different basis functions $\phi_\eta(\bk)$ fall into different representations of crystal symmetry group. 
As a first step, it is instructive to start from isotropic spin-orbit-coupled liquids with the largest symmetry group $SO(3)$,   
 invariant under any arbitrary rotation of space and spin taken in combination. 
Then, the basis functions are labeled by three quantum numbers $\eta=(L, J, J_z)$:  
$L$ is the orbital angular momentum, $J$ is the total angular momentum $J=L+S$ with $S=1$, and $J_z$ is the $z$-component $J$. 
The $2J+1$ basis functions with the same $(L, J)$ and $J_z = -J, -J+1, ...J$ form a multiplet.     
Associated with each $(L, J)$-multiplet is an interaction parameter $F^s_{L, J}$, which parameterizes the excitation energy 
of a particular type of Fermi surface deformation.  
Different $(L, J)$-multiplets correspond to ``orthogonal'' modes of Fermi surface deformations.  

We now explicitly decompose the spin interaction into a few lowest $(L, J)$-multiplets. 
Up to $L=1$, there are four multiplets: $(L=0; J=1)$ and $(L=1; J=0, 1, 2)$, and hence $\delta E_{\rm spin}$ takes the form: 
\begin{widetext}
\beq
\delta E_{\rm spin} = \sum_{\bk, \bk'}\; 
F^s_0 \; {\bf s}(\bk) \cdot {\bf s}(\bk')  
+ F^s_1  \; \left( {\hat \bk} \cdot {\bf s}(\bk) \right) \left( {\hat \bk'} \cdot {\bf s}(\bk') \right)  
+F^s_2  \left( {\hat \bk} \times {\bf s}(\bk)  \right) \cdot \left(  {\hat \bk'} \times {\bf s}(\bk')  \right)   
+  F^s_3 \;  Q_{ij}(\bk) Q_{ij} (\bk') , \label{es}
\eeq
\end{widetext}
where $Q_{ij}=Q_{ji}$ is a second-rank tensor constructed from ${\bk}$ and $\bs(\bk)$:  
\beq
Q_{ij}(\bk) =\frac{1}{2} \left( {\hat k}_i s_j(\bk) + {\hat k}_j s_i(\bk) \right)- \frac{1}{3} \hat{\bk} \cdot \bs(\bk) \delta_{ij}. \label{qij}
\eeq
Eq. (\ref{es}) is a main result of this work, which shows the presence of three $p$-wave ($L=1$) spin interaction channels parameterized by $F^s_{1}, F^s_2$ and $F^s_3$. 
Similar decompositions of spin interaction into higher angular-momentum channels can be carried out by constructing
high-rank tensors from powers of  ${\bk}$ and $\bs(\bk)$. 

At this point,  it is worth discussing the effect of periodic crystal potential, which reduces the full rotational symmetry to its subgroup, the point group of a crystal.     
In this case, Fermi liquid interactions can still be decomposed into different spin-orbit-coupled channels as in (\ref{es}). However, these channels are in one-to-one correspondence 
with the irreducible representations of the point group, instead of the $(L, J)$-multiplets for $SO(3)$ group.  
Despite this difference, for many crystal structures such as cubic, tetragonal, trigonal and hexagonal, 
the four channels in (\ref{es}) remain to be in different point group representations, and hence orthogonal to each other.

When one or more interaction parameters in the spin channel become negative and of sufficiently large magnitude,  Fermi surface instability occurs.  A well-known example is the ferromagnetic instability associated with $F^s_0$ in the $s$-wave spin channel. 
This work is concerned with Fermi liquid instabilities in the $p$-wave (more generally, odd $L$)  interaction channels in spin-orbit-coupled metals. The resulting phases are 
time-reversal-invariant and parity-breaking, and as we will show, exhibit novel properties arising from spin-orbit coupling.   

First, consider the instability associated with $F^s_1$ in the $(L=1; J=0)$ channel. According to (\ref{es}), this instability generates an Ising order parameter: 
\beq
\eta= \sum_{\bk} \hat{\bk} \cdot \bs(\bk). \label{eta}
\eeq
$\eta$ is a pseudo-scalar, because it is invariant under time-reversal and all rotations, but breaks inversion and all reflections. 
Therefore  the ordered phase with $\eta \neq 0$ is an isotropic gyrotropic liquid. 
This gyrotropic order parameter splits the original spin-degenerate Fermi surfaces into two with unequal volumes, with opposite spin polarizations. 
Unlike the case of ferromagnetism, here the spin quantization axis defined in terms of MCBB 
is not uniform but parallel to the momentum: $\bs(\bk) \propto \eta {\hat \bk}$, which leads to a hedgehog spin texture over the Fermi surface.  

Next, consider the instability associated with $F^s_2$ in the $(L=1; J=1)$ channel.  According to Eq.(\ref{es}), this instability generates a vector order parameter 
\beq
\bP = \sum_\bk {\hat \bk} \times    \bs(\bk).    \label{bp}
\eeq
We observe that $\bP$ has the same symmetry as the ferroelectric polarization: it is odd under inversion and invariant under time-reversal, and transforms as a vector under rotation.  
Therefore we identify the ordered phase with $\bP \neq 0$ as a ``ferroelectric'' metal that spontaneously develops a polar axis, despite that its charge polarization is screened by free carriers\cite{anderson}. In this phase, Fermi surfaces are spin-split and deformed by a spontaneously generated spin-orbit field ${\bf h}(\bk)$ 
acting on the original Fermi surface. ${\bf h}(\bk)$ is $\bk$-dependent and proportional to   
the spin polarization field $\bs(\bk)$ generated by the ferroelectric vector order parameter: ${\bf h}(\bk) \propto \bs(\bk)   \propto \bP \times \hat{\bk}$. 
This spin-orbit field has the same form as the Rashba spin-splitting due to an external electric field. 
In our case,  both the Rashba spin splitting of the Fermi surface and the accompanying ferroelectric order $\bP$ are caused by strong electron interactions. 

Lastly, consider the instability associated with $F^s_3$ in the $(L=1, J=2)$ channel. According to (\ref{es}),   
the corresponding order parameter is a traceless symmetric matrix given by  
\beq
Q_{ij} = \sum_{\bk} Q_{ij}(\bk), 
\eeq 
where $Q_{ij}(\bk)$ is defined in (\ref{qij}). This order parameter has $d$-wave symmetry and odd-parity, and hence is a second-rank pseudo-tensor. 
The multipolar phase with $Q_{ij}\neq 0$ can be regarded as an electronic analog of the chiral nematic liquid crystals\cite{lubensky}.    
Its Fermi surfaces are spin-split and deformed by the spin-orbit field $h_i(\bk) \propto s_i(\bk) \propto Q_{ij} k_j $.   
If the matrix $Q_{ij}$ may have two degenerate eigenvalues, the ordered phase is uniaxial; otherwise, it is biaxial.

 To summarize, we find new parity-breaking phases with gyrotropic, ferroelectric and 
multipolar orders in  spin-orbit-coupled Fermi liquids, driven by strong $p$-wave spin interaction.  
Because of spin-orbit coupling, these symmetry-breaking phases exhibit spin-spilt Fermi surfaces. 
The magnitude and direction of the spin-splitting vary strongly over the Fermi surface. 
The characteristic Fermi surface splitting and spin texture in momentum space are predicted to be the hallmark of parity-breaking phases in spin-orbit-coupled metals.  
These features of parity-breaking phases can be detected by angle- and spin-resolved photoemission spectroscopy.  

Since the above parity-breaking order parameters are time reversal invariant and break rotational symmetry of the crystal, they 
couple linearly to lattice distortions that lowers the point group to the same subgroup without enlarging the unit cell. 
As a result, the transition driven by Fermi liquid instability is generally accompanied by a structural transition, from which the electronic order can be inferred.   
A possible exception is the gyrotropic order (\ref{eta}), which preserves the full rotational symmetry of the crystal. 
Such a high degree of symmetry may not be compatible with any lattice distortion caused by atomic displacements. 
In this case, it will be difficult to detect the electronic gyrotropic order with conventional methods\cite{kerr, kerr2, kerr3, kerr4}.

It is important to emphasize that the presence of spin-orbit coupling is indispensable to the parity-breaking phases found in this work.  
To make this point clear, let us consider spin-rotationally-invariant Fermi liquids, whose spin interaction in the $p$-wave channel takes the form $ \sum_{\bk \bk'}  ( {\hat \bk} \cdot {\hat \bk'} ) \; {\bf s}(\bk) \cdot {\bf s}(\bk')$. 
The corresponding Fermi liquid instabilities have been studied in detail\cite{varma, wu1, wu2}. The ordered phases were found to simultaneously 
break two symmetries, the rotational symmetry of space and of spin. As a result, the spin textures are free to rotate as a whole, instead of being rigidly locked 
to momentum as in our case. 
Moreover, these electronic orders cannot couple directly to lattice distortions (which preserve spin rotational symmetry), 
unlike the ferroelectric and multipolar phases of spin-orbit-coupled Fermi liquids.    
We also note that besides spin-orbit coupling, dipolar interactions in ultracold Fermi gases, which also lock spin and momentum,   
can generate ordered phases with similar features\cite{dipolar1,dipolar2}.  In addition, parity-breaking phases can occur in spin-orbit-coupled insulators\cite{motome}.   

Finally, based on recent experiments, we identify several correlated electron materials that show evidence of the above parity-breaking orders. 
First, a recently synthesized material LiOsO$_3$ was found to undergo a second-order ferroelectric structural transition at low temperature\cite{osmate}. 
The high-temperature structure is $D_{3d}$, which is inversion symmetric. A polar axis in $c$ direction 
appears in the low temperature phase, reducing the crystal symmetry to $C_{3v}$.        
Based on the observation of unusually large residual resistivity and Curie-Weiss behavior of spin susceptibility,    
it has been suggested that electron correlation plays an important role and possibly drives the structural transition\cite{osmate}. 
Therefore the low-temperature phase of LiOsO$_3$ may be an electronic-driven ferroelectric metal. 

Second, we suggest pyrochlore oxides A$_2$B$_2$O$_7$ as promising candidates for the multipolar phase. 
The pyrochlore crystal structure has the $O_h$ point group symmetry. 
Due to this crystal anisotropy, the five-component multipolar order parameter $Q_{ij}$ defined in (\ref{qij}) 
splits into a two-dimensional $E_u$ representation $(Q_{xx} - Q_{yy}, 2Q_{zz} - Q_{xx}- Q_{yy})$
and a three-dimensional $T_{2u}$ representation $(Q_{xy}, Q_{yz}, Q_{zx})$. 
Recently, the pyrochlore oxide Cd$_2$Re$_2$O$_7$ was found to undergo a second-order structural transition from cubic to tetragonal at $T_c=200$K, 
with an order parameter  of the $E_u$ symmetry\cite{227-1,227-2,227-3,227-4}. 
Remarkably, the lattice change across the transition is extremely small, whereas    
electrical  properties change drastically.  
In addition, a  large mass enhancement above the transition temperature was inferred from transport and optical measurements\cite{pressure, correlation-1,correlation-2}. 
Therefore,  the structural transition in Cd$_2$Re$_2$O$_7$ may be induced by an electronic transition to the mulitpolar phase.     
      
In the above examples of ferroelectric and multipolar phases, the appearance of electronic order is inferred, by symmetry consideration, from the structural distortion it couples to. 
It is desirable to directly probe the change in electronic structure, Fermi surface, and spin texture across the parity-breaking phase transition via, for example, 
angle- and spin-resolved photoemission spectroscopy. Moreover, it will be interesting to determine whether the driving force for the transition is structural or electronic.     
On the other hand, the isotropic gyrotropic order can be more elusive. In the case of pyrochlore crystals, the gyrotropic order belongs to 
 the $A_{1u}$ representation of the $O_h$ point group, which is incompatible with any phonon mode at the Brillouin zone center\cite{group}. 
 Therefore, it cannot be generated by structural distortions, and if found, should have an electronic origin.   
      
In addition to photoemission, nonlinear optics is a powerful tool for detecting parity-breaking orders described in this work. 
For example, the multipolar phase in tetragonal Cd$_2$Re$_2$O$_7$ has been successfully detected by second-harmonic generation (SHG)\cite{shg}. 
Regarding the gyrotropic order  (\ref{eta}), we find it    
has the same symmetry as the rank-3 isotropic tensor $\epsilon_{ijk}$.  
It then follows from symmetry that this gyrotropic order should lead to sum-frequency generation (SFG)\cite{gyrotropic}, in which   
two incident fields ${\bf E}_{1,2}$ at {\it different} frequencies $\omega_{1,2}$
generate an electric dipole $\bf P$ at the frequency $\omega =  \omega_1 + \omega_2$:  
\beq
P_i (\omega) \propto {\rm sgn}(\eta) \epsilon_{ijk} E_{1, j}(\omega_1) E_{2, k}(\omega_2). 
\eeq
This nonlinear optical effect gives a direct way of detecting the much-hidden gyrotropic order. 

Our work on spin-orbit coupled metals leaves a number of open questions for future studies. It is worthwhile to relate the phenomenological parameters in Fermi liquid theory to microscopic 
interactions. It will be extremely interesting to study superconducting instabilities of spin-orbit-coupled Fermi liquids, especially those proximate to the parity-breaking phases.

{\it Note:} A recent work shows that optical circular dichroism can be used to probe the parity-breaking phases  described in this work\cite{norman}.

\begin{acknowledgments}
I thank Tim Hsieh and Vlad Kozii for interesting discussions. This work is supported by David and Lucile Packard Foundation.  
\end{acknowledgments} 

\bibliographystyle{apsrev}

\newpage

\end{document}